\documentclass{iopart}
\usepackage{iopams}  
\usepackage{graphicx}
\newcommand{\bm}[1]{\mbox{\boldmath$#1$}}
\newcommand{\be}{\begin{equation}}
\newcommand{\ee}{\end{equation}}
\newcommand{\bea}{\begin{eqnarray}}
\newcommand{\eea}{\end{eqnarray}}
\begin{document}

\title[Diffusion of a ring polymer via the Brownian dynamics]
{Diffusion of a ring polymer in good solution via the Brownian dynamics 
with no bond crossing}

\author{Naoko Kanaeda and Tetsuo Deguchi}
\address{Department of Physics, Ochanomizu University, Tokyo 112-8610, Japan}
\eads{\mailto{kanaeda@degway.phys.ocha.ac.jp}, \mailto{deguchi@phys.ocha.ac.jp}}
\date{\today}

\begin{abstract}  Diffusion constants 
$D_{R}$ and $D_{L}$ of ring and linear polymers 
of the same molecular weight in a good solvent, respectively,  
have been evaluated through the Brownian dynamics 
with hydrodynamic interaction in which  no bond crossing is possible. 
The ratio $C=D_{R}/D_{L}$, which should be universal in the context of  
the renormalization group, 
has been estimated as $C= 1.14 \pm 0.01$ for the large-$N$ limit.  
It should be consistent with that of synthetic polymers, 
while it is smaller than that of DNAs such as $C \approx 1.3$. 
We also perform the same simulation through Brownian dynamics 
with hydrodynamic interaction where bond crossings are possible,  
and obtain almost the same estimate for the ratio $C$.  
\end{abstract}
%
%
%
%
%

\maketitle

\section{Introduction} 

Recently, there has been much progress in 
experimental techniques associated with ring polymers.  
Ring polymers of large molecular weights 
are synthesized not only quite effectively \cite{Bielawski}  
but also with small dispersions and high purity \cite{Takano05,Takano07}. 
Diffusion constants of linear, 
relaxed circular and supercoiled DNAs have been measured quite 
accurately \cite{PNAS06}. Furthermore, hydrodynamic radius of circular DNA 
has also been measured \cite{Araki06}.  
The developments are quite remarkable. In fact,  
it used to be considered quite difficult to synthesize 
ring polymers of large molecular weights.  
It has now become quite interesting to evaluate numerically 
dynamical or conformational quantities  of linear and ring polymers 
that can be measured in experiments.

It should be nontrivial  
how linear and ring polymers with the same 
molecular weight in solution may have 
different dynamical or conformational properties. 
In fact, the excluded volume effect should play a more significant role  
for ring polymers than for linear polymers, 
since the average distance among monomers 
is smaller due to the constraint of closing two ends \cite{Cala02}. 
Moreover, in a dilute solution, 
the topology of a given ring polymer is 
conserved under thermal fluctuations \cite{Whitt07}
and represented by a knot. 
Topological constraints may lead to nontrivial statistical mechanical 
or dynamical properties of 
ring polymers \cite{Vologodskii74,Muthukumar,Quake94,Univ97,GrosbergPRL00,Lai02,Shima02,Akos03,Zonta05,Orlandini07}.

In the paper we discuss diffusion constants $D_R$ and $D_L$
of ring and linear polymers in good solution, respectively, 
 via the Brownian dynamics with hydrodynamic interaction in which  
bond crossings are effectively prohibited. 
Here the ring and linear polymers have the same molecular weight, and we 
calculate diffusion constants for several different values of 
the number of segments, $N$, for $5 < N < 50$.  
We then calculate the ratio  $C= D_R/D_L$ 
and compare it with the values measured in some experiments and 
other theoretical values. 
This gives a test for the validity of dynamical models 
of ring and linear polymers. In fact, it is suggested  from  
the renormalization group argument that the large-$N$ limit of 
$C$ should be universal among some class of polymer models.   
Hereafter we call the Brownian dynamics with no bond crossings 
{\it dynamics A}. 

Furthermore, we also perform the Brownian dynamics with hydrodynamic 
interaction under almost the same molecular potentials as dynamics A except 
the parameters of the FENE (finite extensible non-linear elongational) 
potential which determine the maximal distance between neighboring 
monomers. We set the maximal distance 
larger so that bond crossings are allowed. 
We call it {\it dynamics B}. 
Dynamics B has precisely the same potential parameters as  
that of Ref. \cite{Rey}. 
We have found that bond crossings occurred for dynamics B, 
checking the topology of the ring polymer by 
calculating some knot invariants at every time step 
of the Brownian dynamics. 

Simulation results of both dynamics A and B should be important. 
In fact, there have been several simulation results obtained and 
accumulated for dynamics B \cite{Rey,Cifre}. 
We may compare the present simulation 
with previous ones. In this sense,  dynamics B is a   
standard algorithm in the Brownian dynamics. Furthermore, 
dynamics A is important since it preserves 
the initial topology of a ring polymer.  

The present study should be useful for making explicit 
connections between experimental and theoretical results of 
dilute solutions of ring polymers.
In fact, for dilute ring-polymer solutions, even some fundamental properties 
such as the effects of topological constraints have not been clearly 
confirmed in experiments, yet. 
Through simulations, we can study the effects of topological constraints, 
 which can be checked in  experiments.

The content of the paper consists 
of the following: In section 2, we briefly explain the simulation method.  
In section 3, we discuss three simulation results. In subsection 3.1 
 the ratio of the mean square 
radii of gyration of ring and linear polymers, $g$,   
are evaluated numerically.    
The value of $g$ for dynamics B is 
consistent with the lattice simulation result, 
while that of dynamics A is larger than the standard one. We confirm it also 
by the Monte-Carlo simulation. 
It should thus be an interesting future problem to evaluate 
the ratio $g$ for  larger values of $N$. 
In subsection 3.2, we discuss the ratio $C$  
both for dynamics A and B. We find that the estimates of $C$ are given by 
almost the same value both for dynamics A and B. Interestingly, 
the estimate of $C$ is consistent with a theoretical value 
given by a perturbation theory, while it is different 
from that of the renormalization group calculation in one-loop order. 
However, we should note that a one-loop order evaluation  
could give only a rough estimate and 
multi-loop corrections could improve it.

\section{Simulation method}

The ring polymer molecule is modeled as a cyclic bead-and-spring chain 
with $N$ beads connected by $N$ FENE 
(finite extensible non-linear elongational) springs 
with the following force law:  
\be 
{\bm F}({\bm r})=- H {\bm r}/(1-r^2/r^2_{max}) \, , 
\ee 
where $r=|{\bm r}|$. Let us denote by $b$ the unit of distance. Here 
we assume that the average distance  between neighboring monomers 
is  approximately given by $b$.  
We set constants $H$ and $r_{max}$ as follows:  
$H=30 k_B T/b^2$ and $r_{max}=1.3 b$ for dynamics A, and 
$H=3 k_B T/b^2$  and $r_{max}=10 b$ for dynamics B. 
We assume the Lennard-Jones (LJ) potential acting among monomers as follows. 
\be 
V(r) = 4 \epsilon_{\rm LJ} \left[ \left( {\frac {\sigma_{\rm LJ}} {r_{ij}} 
} \right)^{12} 
- \left({\frac {\sigma_{\rm LJ}} {r_{ij}} } \right)^{6}   \right] 
\ee
Here $r_{ij}$ is the distance of beads $i$ and $j$, and $\epsilon_{\rm LJ}$  
and $\sigma_{\rm LJ}$ denote the minimum energy and the zero energy distance, 
respectively \cite{Cifre}. 
We set the Lennard-Jones parameters 
as $\sigma_{LJ}=0.8 b$ and $\epsilon_{LJ}=0.1 k_B T$ 
so that they give good solvent conditions \cite{Rey}.    
Here $k_B$ denotes the Boltzmann constant. 

We employ the predictor-corrector version \cite{Iniesta} 
of the Ermak-McCammon algorithm \cite{Ermak} 
for generating the time evolution of 
a ring polymer in solution. The details are given in Appendix A. 
The hydrodynamic interaction is taken into account 
through the Ronte-Prager-Yamakawa tensor \cite{Rotne,Yamakawa}   
where the bead friction is given by $\zeta=6 \pi \eta_s a$ 
with the bead radius $a=0.257b$ and a dimensionless 
hydrodynamic interaction parameter 
$h^{*}=(\zeta/6 \pi \eta_s) \sqrt{H/\pi k_B T}= 0.25$.  

In the present simulation, physical quantities are given in dimensionless 
units such as in Ref. \cite{Cifre}.  We divide length by $b$, 
energy by $k_B T$ and time by $\zeta b^2/k_B T$. 
Let us indicate dimensionless quantities by an asterisk as superscript.  
We have $H^{*} = 30$, $r^{*}_{max}=1.3$ for dynamics A, and 
$H^{*} = 3$, $r^{*}_{max}=10$ for dynamics B. 
We take the simulation time step      
$\Delta t^{*}= 10 ^{-4}$. 

When we evaluate the mean square radius of gyration and the  
diffusion constant for ring and linear polymers 
through the Brownian dynamics with hydrodynamic interaction, 
we keep each run long enough so that the diffusion constant approaches 
its equilibrium value.  
For instance, in the case of linear polymers of $N=45$ of dynamics A,   
we have performed $9.4 \times 10^{5}$ time steps for each run. 
After the average value of the diffusion constant approaches 
some equilibrium value,  we start sampling the data and pick up 
one conformation out of every 18, 800 time steps.   
Then, the diffusion constant evaluated  
at 740,000th time step is given by $9.256 \times 10^{-2}$, 
 while that at 940,000 th time step is given by 
$9.276 \times 10^{-2}$. The difference $0.020 \times 10^{-2}$
 is smaller than their probable error $0.068 \times 10^{-2}$.

\section{Simulation results}
\subsection{Ratio of the mean square radii of gyration}

The mean square radius of gyration $\langle R_G^2 \rangle$ 
of a polymer consisting of $N$ monomers 
is defined by 
$$
\langle R_G^2 \rangle = {\frac 1 N}  
\sum_{j=1}^{N} \langle ({\vec r}_j - {\vec r}_G)^2 \rangle 
$$
where ${\vec r}_j$ denote the position vectors of monomers 
for $j=1, 2, \ldots, N$, and ${\vec r}_G$ the position vector 
of the center of mass of the polymer.  The symbol $\langle A \rangle$ 
denotes the ensemble average of physical quantity $A$. 

Let us discuss the estimates of the mean square radius of gyration 
for ring and linear polymers, $\langle R^2_G \rangle_R$ 
and $\langle R^2_G \rangle_L$, respectively,  
obtained by dynamics A and dynamics B. 
They are  plotted in Figure 1 against the number of segments $N$ 
in the double logarithmic scales.  
It is clear that they are fitted well by straight lines. It seems that  
the $N$-dependence is close to that of the asymptotic behavior,  
although the number of segments $N$ are not very large, yet.  
As we shall discuss later, it is probably due to the effect  
of the off-lattice molecular potentials 
employed in the dynamics. Thus,    
as a fitting formula, we employ the large-$N$ asymptotic behavior 
of the mean square radius of gyration: 
$\langle R^2_G \rangle = A N^{2 \nu}$.  
The estimates of the fitting parameters, 
$A_R$ and $\nu_R$ for ring polymers, and 
$A_L$ and $\nu_L$ for linear polymers, 
are given in the caption of Figure 1.

Let us now define the geometric shrinking factor $g$  by \cite{Burchard}
\begin{equation} 
g = \langle R_G^2 \rangle_R / \langle R_G^2 \rangle_L . 
\end{equation} 
We assume that exponent $\nu$ should be the same 
for ring and linear chains, i.e. $\nu_R=\nu_L$.   
We thus have the following fitting formula with three parameters:  
\begin{equation} 
g=g_{\infty} \left( 1 + B_g N^{-\Delta_g} \right) . \label{eq:g-fit} 
\end{equation}  
Applying (\ref{eq:g-fit}),  
we have $g_{\infty} = 0.559 \pm 0.007$ 
for dynamics A and $g_{\infty} = 0.535 \pm 0.002$ for 
dynamics B, as shown in Figure 2.

The estimate of $g$ value for
dynamics B, $g_{\infty}=0.535 \pm 0.002$, 
should be  consistent with the Monte Carlo simulation 
using the bond fluctuation model \cite{Ziff01}. 
Interestingly, however, 
the estimate of $g_{\infty}$ for dynamics A is larger than 
that of dynamics B even if we take into account their errors. 
The enhancement of value $g_{\infty}$ in dynamics A 
should be due to the potential forces.  
In fact, we have 
confirmed that almost the same value of $g_{\infty}$ is obtained 
by the Monte-Carlo simulation with the same 
molecular potentials as dynamics A. 
Therefore, we conclude that 
it is due to the potential forces employed in dynamics A. 
Here, the potential function of the Monte-Carlo simulation 
of linear chains is given by the following. 
$$
-\sum_{i=1}^{N-1} 0.5 H \, r^2_{max} \, \ln[1-(r_{i,i+1}/r_{max})^2]
+4 \epsilon_{\rm LJ} \sum_{i>j}^{N}
[(\sigma_{\rm LJ}/r_{ij})^{12}-(\sigma_{\rm LJ}/r_{ij})^{6}] \, . 
$$
For ring chains, we add a term of $r_{N, 1}$ due to the periodicity.  

Here we note that we have employed the symbol 
$g_{\infty} $ for the fitting parameter, expecting that 
it should suggest the asymptotic value of $g$.   
However, in order to evaluate the true asymptotic value of $g$,  
we have to perform simulations for larger values of $N$.  
It should thus be an interesting future problem 
whether the enhancement of value $g$ should be relevant to the 
asymptotic value of $g$ or not.

\begin{figure}
  \begin{center}
    \begin{tabular}{cc}
  \resizebox{160mm}{!}{\includegraphics{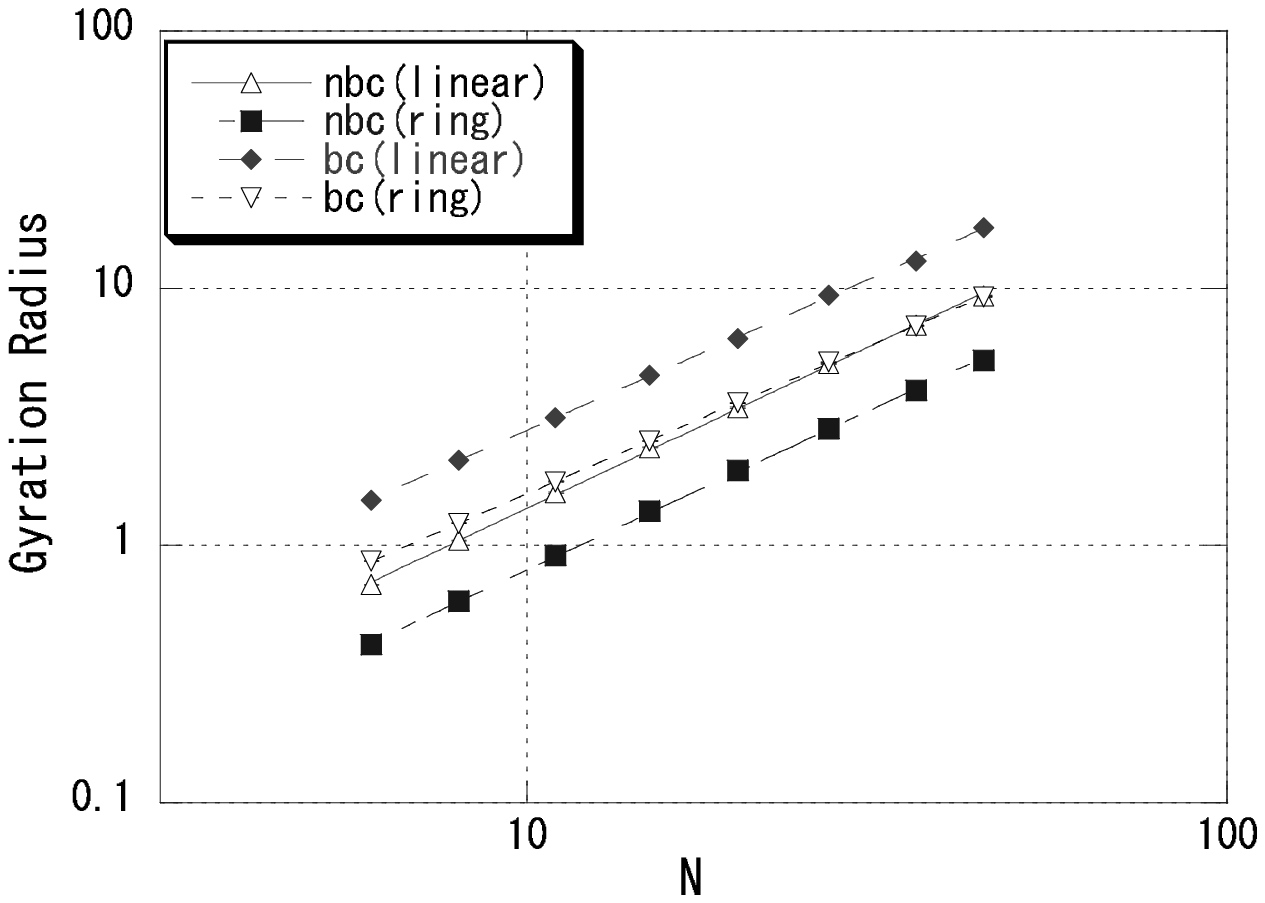}}
    \end{tabular}
    \label{fig:gyration_radius}
  \end{center}
\caption{Mean square radius of gyration $\langle R_G^2 \rangle_L$ and 
$\langle R_G^2 \rangle_R$ for dynamics A (
no bond crossing, nbc) and B (allowed bond crossings, bc). 
For dynamics A, data points of linear polymers are shown by 
\opentriangle,  where 
ring polymers by \fullsquare, 
where $A_R=0.043 \pm 0.0001$ and $\nu_R= 1.270 \pm 0.001$.
For dynamics B, linear polymers 
by \fulldiamond, where $A_L=0.175 \pm 0.001$ and $\nu_L= 1.204 \pm 0.002$;  
ring polymers by \opentriangledown, 
where $A_R=0.105 \pm 0.001$ and $\nu_R= 1.179 \pm 0.001$.   
}
\end{figure}

According to the one-loop renormalization group calculation 
\cite{Prentice} $g$ is given by 
\be 
g_{\infty} = \exp(13/96)/2 = 0.573. \label{eq:g-rg}
\ee
The value (\ref{eq:g-rg}) is  larger 
than the estimates,
$g_{\infty} = 0.559 \pm 0.007$ 
for dynamics A and $g_{\infty} = 0.535 \pm 0.002$ for dynamics B.  
Thus, the one-loop calculation does not explain the 
estimate of $g_{\infty}$ for dynamics A or B. 
However, we should note that it is possible that 
the one-loop RG result gives only a crude approximation,  
and higher-order calculation improves the $g$ value.   
Here we note that for the $\epsilon$-expansion of the $n$-vector model, 
 higher-order terms have been calculated 
in order to evaluate universal quantities \cite{Zinn-Justin}.  
Thus, the multi-loop corrections should be important, 
although the one-loop correction \cite{Prentice} 
is based on Fixman's cluster expansion 
\cite{Oono} and it is not clear whether one can extend it.

Through perturbation calculation, 
$g$ was estimated in terms of the excluded-volume parameter $z$ 
as  follows \cite{Zimm-Stockmayer,Casassa}: 
\be 
g = {\frac 1 2}
\left[ 1+ \left({\frac {\pi} 2} - {\frac {134} {105}} \right) z 
+ \cdots \right] . 
\ee
The value of $g$ is dependent on the excluded-volume parameter, $z$.  
In order to have $g \approx 0.53$, we have $z \approx 0.20$.  
Here we note that the value $z$ depends on the number of segments $N$. 
In order to have  $z \approx 0.20$ 
we have to adjust many model parameters. 
Thus, it should be practically impossible to give good estimates of 
$g$ by making use of the perturbation theory.

\begin{figure}
  \begin{center}
    \begin{tabular}{cc}
  \resizebox{160mm}{!}{\includegraphics{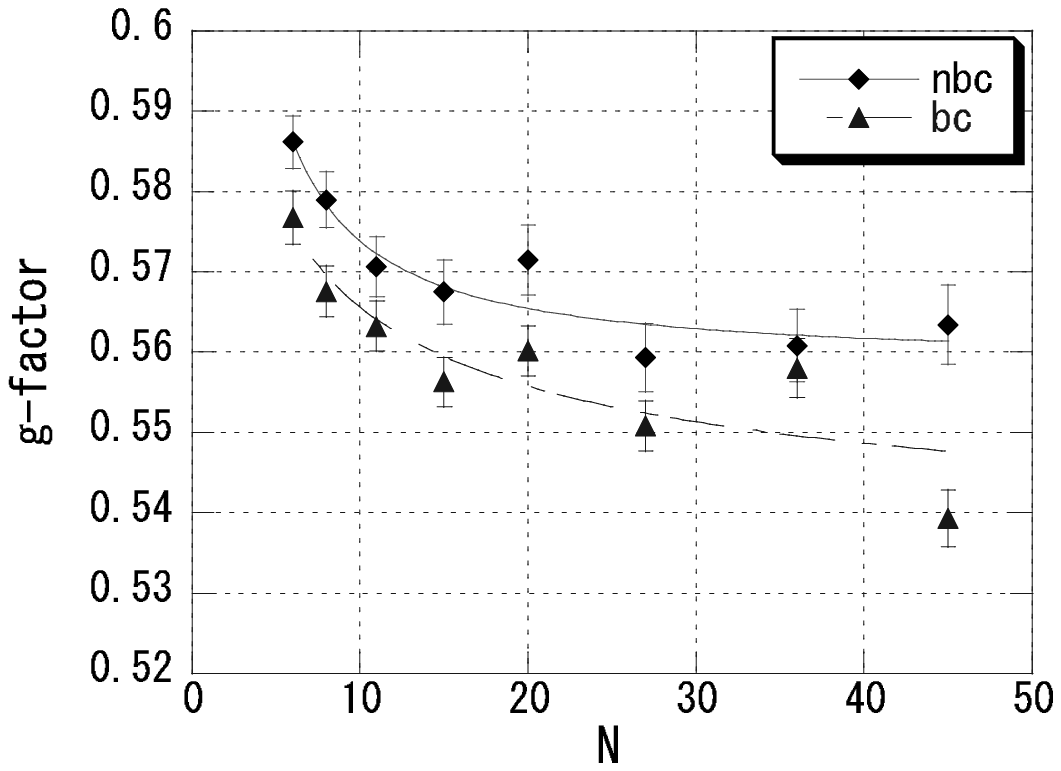}}
    \end{tabular}
    \label{fig:g-factor}
  \end{center}
\caption{The ratio 
$g= \langle R_G^2 \rangle_R / \langle R_G^2 \rangle_L$ versus $N$ with 
the fitting curve (\ref{eq:g-fit}). For dynamics A 
shown by \fulldiamond (nbc), we have   
$g_{\infty} = 0.559 \pm 0.007$,   
$B_g=0.402 \pm 0.438$ and $\Delta_g=1.173 \pm 0.706$. 
Here $\chi^2=3.3$ for 8 data points.  
For dynamics B shown by \fulltriangle (bc), we have   
$g_{\infty} = 0.535 \pm 0.002$  
$B_g=0.204 \pm 0.079$ and $\Delta_g=0.565 \pm 0.476$. 
Here $\chi^2=14.9$ for 8 data points. 
}
\end{figure}

\begin{table}
\begin{center}
\begin{tabular}{cccc}
\hline
N& $\langle R_G^2 \rangle_R$&$\langle R_G^2 \rangle_L$ & 
$g={\langle R_G^2 \rangle_R}/{\langle R_G^2 \rangle_L}$\\
\hline
$6$
&$0.413\pm0.001$ &$0.705\pm0.003$ & $0.586\pm0.003$ \\
$8$
&$0.606\pm0.001$ &$1.046\pm0.004$ & $0.579\pm0.003$ \\
$11$
&$0.916\pm0.002$ &$1.605\pm0.007$ & $0.571\pm0.004$ \\
$15$ 
&$1.364\pm0.002$ &$2.404\pm0.012$ & $0.567\pm0.004$ \\
$20$
&$1.961\pm0.005$ &$3.431\pm0.018$ & $0.571\pm0.004$ \\
$27$
&$2.842\pm0.008$ &$5.082\pm0.024$ & $0.559\pm0.004$ \\
$36$
&$4.028\pm0.011$ &$7.182\pm0.037$ & $0.561\pm0.004$ \\
$45$
&$5.260\pm0.016$ &$9.335\pm0.053$ & $0.563\pm0.005$ \\
\hline
\end{tabular}
\caption{Dynamics A (no bond crossing).  
Mean square radii of gyration for linear and ring polymers, 
$\langle R_G^2 \rangle_R$ and $\langle R_G^2 \rangle_L$, and 
the $g$ values. Applying  
the least square method for 
$ \langle R_G^2 \rangle_L = A_L N^{2\nu_L}$ and  
$ \langle R_G^2 \rangle_R = A_R N^{2\nu_R}$, respectively, 
the following estimates are obtained:  
$2\nu_L= 1.288 \pm 0.002, A_L=0.072 \pm 0.001;$ 
$2\nu_R = 1.270 \pm 0.001, A_R=0.043 \pm 0.001$. 
In all the tables, errors are given by probable errors. 
}
\end{center}
\end{table}

\begin{table}
\begin{center}
\begin{tabular}{cccc}
\hline
N& $\langle R_G^2 \rangle_R$&$\langle R_G^2 \rangle_L$ & 
$g={\langle R_G^2 \rangle_R}/{\langle R_G^2 \rangle_L}$\\
\hline
$6$
&$0.868\pm0.002$ &$1.505\pm0.005$ & $0.577\pm0.003$ \\
$8$
&$1.221\pm0.003$ &$2.151\pm0.008$ & $0.568\pm0.003$ \\
$11$
&$1.774\pm0.004$ &$3.149\pm0.011$ & $0.563\pm0.003$ \\
$15$ 
&$2.550\pm0.005$ &$4.583\pm0.016$ & $0.556\pm0.003$ \\
$20$
&$3.587\pm0.008$ &$6.404\pm0.022$ & $0.560\pm0.003$ \\
$27$
&$5.172\pm0.012$ &$9.389\pm0.037$ & $0.551\pm0.003$ \\
$36$
&$7.155\pm0.017$ &$12.821\pm0.054$ & $0.558\pm0.004$ \\
$45$
&$9.306\pm0.023$ &$17.264\pm0.070$ & $0.539\pm0.004$ \\
\hline
\end{tabular}
\caption{Dynamics B (allowed bond crossings). 
Mean square radii of gyration for linear and ring polymers, 
$\langle R_G^2 \rangle_R$ and $\langle R_G^2 \rangle_L$, and 
the $g$ values. 
Applying  the least square method for   
$ \langle R_G^2 \rangle_L = A_L N^{2\nu_L}$ and  
$ \langle R_G^2 \rangle_R = A_R N^{2\nu_R}$, respectively, 
the following estimates are obtained:  
$2\nu_L= 1.204 \pm 0.002, A_L=0.175 \pm 0.001$;  
$2\nu_R = 1.179 \pm 0.001, A_R=0.105 \pm 0.001$. 
}
\end{center}
\end{table}

\subsection{Ratio of diffusion constants}

\begin{figure}
  \begin{center}
    \begin{tabular}{cc} 
{\resizebox{160mm}{!}{\includegraphics{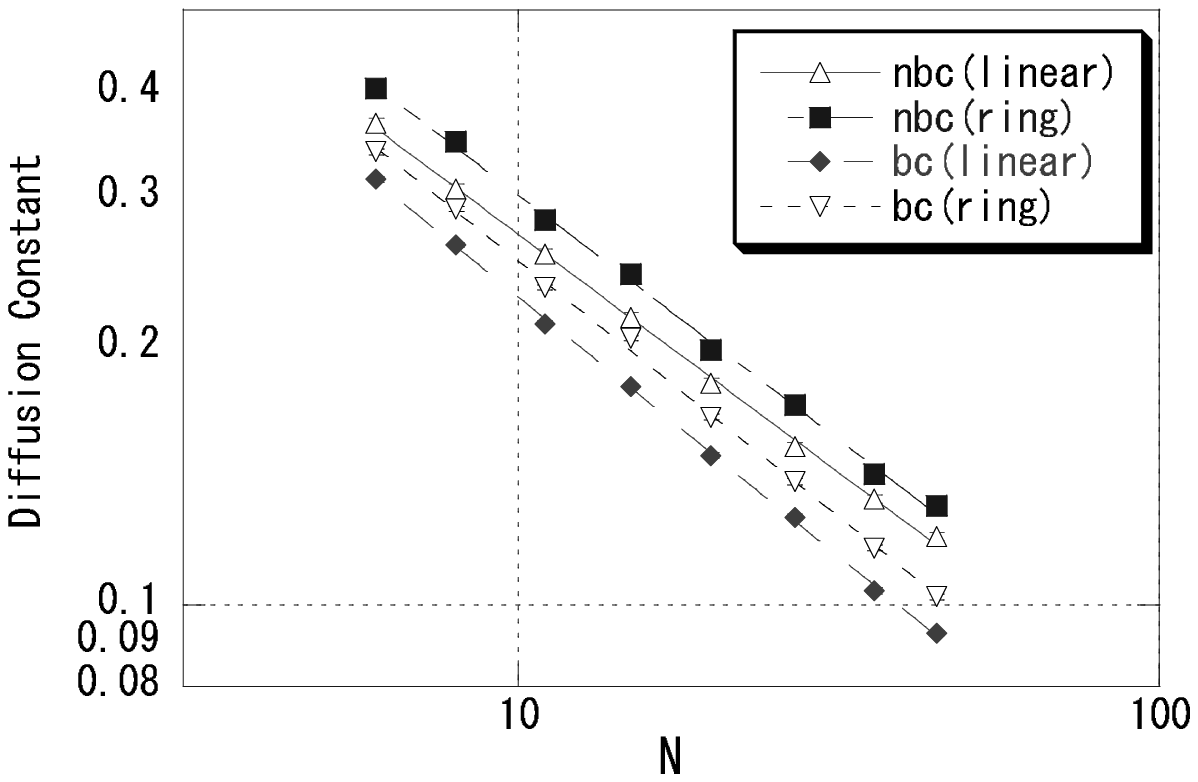}}}
    \end{tabular}
    \label{fig:diffusion-constants}
  \end{center}
\caption{Diffusion constants of ring and linear polymers for dynamics A (nbc) 
depicted by \opentriangle  and \fullsquare, respectively, 
for dynamics B (bc) 
\fulldiamond  and \opentriangledown, respectively. 
The horizontal axis denotes 
the number of segments, $N$. }
\end{figure}
 
Let us recall that the diffusion constant of a polymer 
is defined by the following: 
\be 
D =  \lim_{t \rightarrow \infty} {\frac 1 {6t}} 
\langle (\vec{r}_G(t)-\vec{r}_G(0))^2  \rangle 
\label{eq:dfD}
\ee
Here ${\vec r}_G(t)$ denote the position vector of the center of mass 
of the polymer. 
Making use of (\ref{eq:dfD}) 
we have evaluated the diffusion constant of ring and linear polymers, 
$D_{R}$ and $D_{L}$, respectively, through dynamics A and B. 

According to the Einstein relation, the diffusion constant of 
a polymer should be given by $D = k_B T / \zeta$ where $\zeta$ is given by 
$\zeta= 6 \pi \eta R_H$ with viscosity $\eta$ and the hydrodynamic 
radius $R_H$.  
Let us assume that the hydrodynamic radius 
$R_H$ has the same asymptotic scaling behavior with the square root 
of the mean square radius of gyration:   
$\sqrt{\langle R^2_G \rangle} \propto N^{\nu}$. 
Thus, in a dilute solution,  
we have the following large-$N$ behavior: 
\begin{equation}
D = {\frac {k_B T} {6 \pi \eta R_H} 
}  
 \propto  N^{-\nu} \, . 
\label{eq:asympt-D}
\end{equation}
Taking the analogy of the large-$N$ behavior (\ref{eq:asympt-D}), 
we introduce the following fitting formulas: 
$D_R = A(D_R) N^{-\nu(D_R)}$ and $D_L = A(D_L) N^{-\nu(D_L)}$.  
Applying them to the data of Table 2, we have the estimates as shown 
in the caption of Figure 3. 
The fitting curves are shown in Figure 3.  
The estimates of $\nu(D_R)$ and $\nu(D_L)$ 
are consistent with the expected 
$N$-dependence: $D_R$, $D_L \propto N^{- \nu}$ with $\nu \approx 0.59$.

Thus, formula (\ref{eq:asympt-D}) gives good fitting curves 
to the graphs of the diffusion constants $D_R$ and $D_L$ versus $N$, 
and the estimates of the exponents $\nu(D_R)$ and $\nu(D_L)$  
are at least roughly in agreement with 
the SAW exponent $\nu_{\rm SAW}=0.588$,   
although the large-$N$ behavior (\ref{eq:asympt-D}) 
should be valid only when $N$ is asymptotically large enough. 
It is likely that  $N=50$ is not large enough to investigate 
any asymptotic behavior of the diffusion constants.

\begin{table}
\begin{center}
\begin{tabular}{cccc} \hline
N & $D_R$ & $D_L$ & $C={D_R}/{D_L}$ \\ \hline
$6$ & $0.404\pm0.005$ & $0.368\pm0.005$ & $1.099\pm0.028$ \\
$8$ & $0.350\pm0.004$ & $0.308\pm0.004$ & $1.137\pm0.029$  \\
$11$ & $0.284\pm0.004$ & $0.258\pm0.003$ & $1.098\pm0.028$  \\
$15$ & $0.244\pm0.003$&$0.218\pm0.003$ &$1.123\pm0.028$ \\
$20$&$0.199\pm0.003$&$0.182\pm0.002$ &$1.095\pm0.027$  \\
$27$&$0.172\pm0.002$&$0.152\pm0.002$ &$1.120\pm0.028$  \\
$36$&$0.142\pm0.002$&$0.133\pm0.002$ &$1.071\pm0.024$  \\
$45$&$0.131\pm0.002$&$0.120\pm0.001$ &$1.087\pm0.026$  \\
\hline
\end{tabular}
\caption{Dynamics A (no bond crossing): 
Diffusion constants of ring and linear polymers, 
$D_R$ and $D_L$, and the $C$ values.  Each estimate 
is derived from the average over more than 2,000 runs.
We have $A(D_R)=1.011 \pm 0.012, \nu(D_R)=0.601 \pm 0.004$; 
$A(D_L)=0.938 \pm 0.107,  \nu(D_L)=0.610 \pm 0.004$.
 }
\end{center}
\end{table}

\begin{table}
\begin{center}
\begin{tabular}{cccc} \hline
N & $D_R$ & $D_L$ & $C={D_R}/{D_L}$ \\ \hline
$6$ & $0.341\pm0.003$ & $0.316\pm0.002$ & $1.078\pm0.016$ \\
$8$ & $0.292\pm0.002$ & $0.265\pm0.002$ & $1.102\pm0.016$  \\
$11$ & $0.236\pm0.002$ & $0.214\pm0.002$ & $1.104\pm0.016$  \\
$15$ & $0.206\pm0.001$&$0.180\pm0.001$ &$1.140\pm0.016$ \\
$20$&$0.166\pm0.001$&$0.150\pm0.001$ &$1.109\pm0.016$  \\
$27$&$0.139\pm0.001$&$0.127\pm0.001$ &$1.100\pm0.017$  \\
$36$&$0.117\pm0.001$&$0.104\pm0.001$ &$1.122\pm0.018$  \\
$45$&$0.102\pm0.001$&$0.093\pm0.001$ &$1.107\pm0.017$  \\
\hline
\end{tabular}
\caption{Dynamics B (allowed bond crossings): 
Diffusion constants of ring and linear polymers, 
$D_R$ and $D_L$, and the $C$ values.  Each estimate 
is given by the average over more than 4,000 runs.
We have $A(D_R)=1.138 \pm 0.022, \nu(D_R)=0.575 \pm 0.007$; 
$A(D_L)=1.138 \pm 0.022,  \nu(D_L)=0.610 \pm 0.004$.   
}
\end{center}
\end{table}

It is clear from Tables 3 and 4 that 
the estimates of $C$ are almost the same for dynamics A and B. Here,  
in Figure 4, the $C$ values are plotted against the number of 
segments $N$ for dynamics A and B, respectively. 
We also observe that    the estimates of $C$ 
are independent of the number of segments, $N$. In fact, 
it is also the case with the experimental results of 
DNAs \cite{PNAS06}.

Let us assume again that exponent $\nu$ should be the same 
for the diffusion constants of ring and linear chains, $D_R$ and 
$D_L$, respectively.   
Applying the fitting formula 
\begin{equation}
C=C_{\infty} \left( 1 + B_C N^{-\Delta_C} \right), \label{eq:C-fit} 
\end{equation}  
we obtain the following estimate: 
$C_{\infty} = 1.14 \pm 0.01$ for dynamics A; 
$C_{\infty} = 1.11 \pm 0.01$ for dynamics B. 
The fitting curves are shown in Figure 4.

According to the one-loop renormalization group calculation 
in the presence of both hydrodynamic and self-avoiding interactions 
\cite{Oono-Kohmoto,Schaub}, a universal ratio $C$ is given by 
\be 
C_{\infty} \equiv \lim_{N \rightarrow \infty} D_R/D_{L} = \exp(3/8) 
= 1.454.  \label{eq:C-rg}
\ee
The value (\ref{eq:C-rg}) is much larger 
than the estimate of $C= 1.14 \pm 0.01$ for dynamics A 
and $C= 1.11 \pm 0.01$ for dynamics B. 
As in the case of the $g$ value, it is possible that 
the one-loop order result gives only a crude result. 
Thus, higher-order RG corrections should be important.  
Here we note that the one-loop calculation was performed through 
the conformation-space renormalization-group approach \cite{Oono}, 
and it would be nontrivial to calculate higher order corrections.

Some years ago, the ratio $C$ has been estimated by 
the perturbative calculation 
in terms of the excluded-volume parameter $z$ \cite{Fukatsu}: 
\be 
C = D_R/D_L = {\frac {3 \pi} 8}
\left( \frac {1+ 1.827 z} {1 + 1.890 z} \right)^{1/3} .   
\label{eq:C-value}
\ee
The value of $C$ is rather constant with respect to $z$.  
We have 1.178 at $z=0$, and 1.165 at $z=\infty$.  
It is interesting to note that the theoretical value 
(\ref{eq:C-value}) is rather close to 
the simulation value,  $C=1.14 \pm 0.01$. 
Thus, the perturbative calculation gives 
a theoretical value consistent with the simulation result
although the validity of the perturbation theory is not clear.

Diffusion constants $D_R$ and $D_L$ 
have been measured in several experiments. 
We observe a tendency that for synthetic polymers 
$C$ is given by 1.1 to 1.2, while 
for linear and circular DNAs it is roughly given by 
1.3. For instance, it is estimated for relaxed circular DNAs  
as $C= 1.32 \pm 0.014$ \cite{PNAS06}.   
For synthetic polymers through scattering experiments, 
$C=1.1 \sim 1.2$ \cite{Duval85} and  $C=1.07 \sim 1.15$ \cite{Hodgson91}.  
Here we note that in Ref. \cite{Griffiths95} 
 $C$ is estimated as a little larger value 
than in other synthetic polymer experiments. 

We thus conclude that the present model of ring and linear polymers 
should be valid for synthetic polymers, while for relaxed circular DNAs 
some additional potential energy might be important.

\begin{figure}
  \begin{center}
    \begin{tabular}{cc}
      \resizebox{150mm}{!}
{\includegraphics{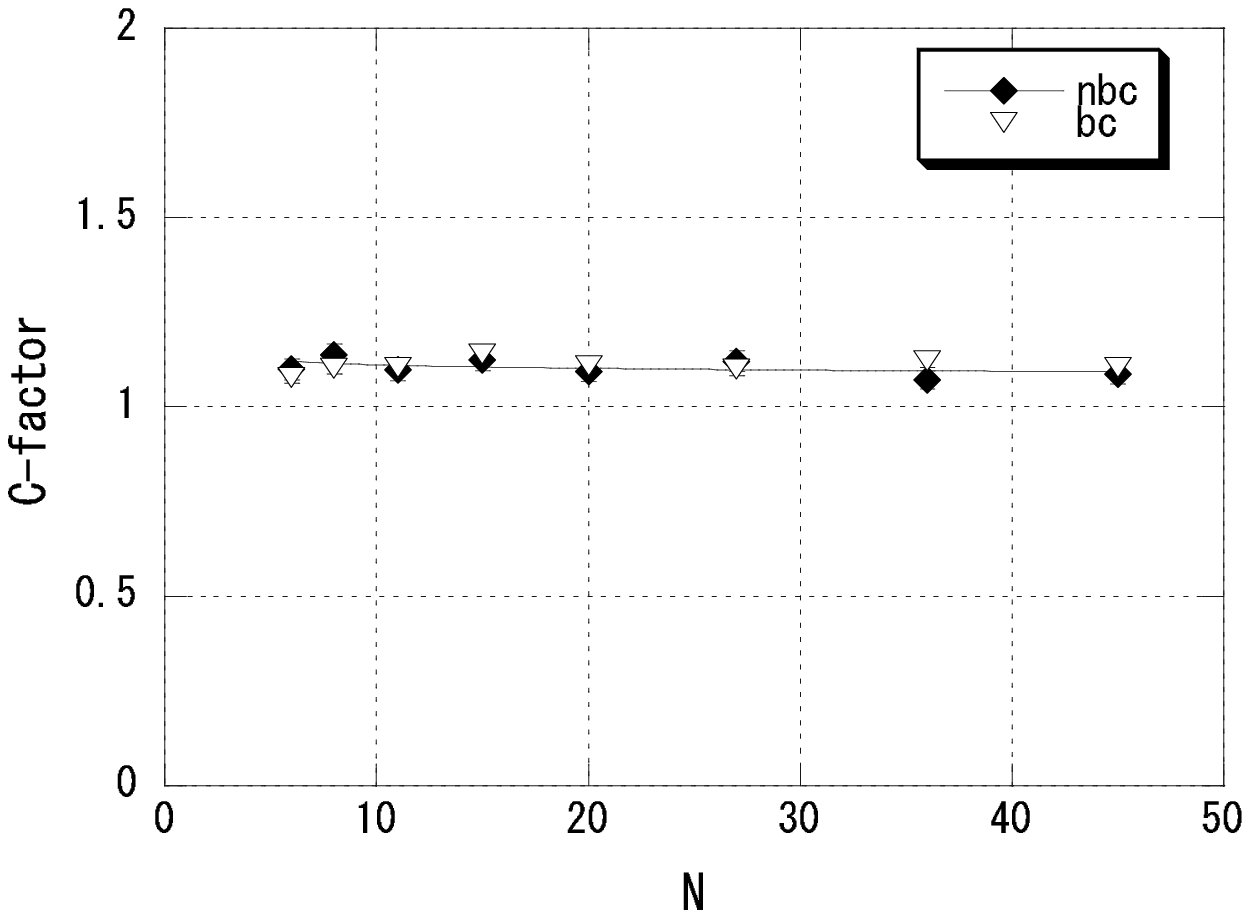}} 
    \end{tabular}
    \label{fig:C-value-B}
  \end{center}
\caption{
Ratio $C=D_R/D_L$ versus $N$.  
For dynamics A (\fulldiamond) with fitting curve 
(\ref{eq:C-fit}),  
$C_{\infty} = 1.14 \pm 0.01$,  
$B_C= 0.094 \pm 76.4$ and $\Delta_C =2.98 \pm 454.97$. 
Here  $\chi^2=6.0$ for 8 data points. For dynamics B 
(\opentriangledown),  $C_{\infty} = 1.11 \pm 0.01$,  
$B_C= 161.8 \pm 2010.3$ and $\Delta_C =4.78 \pm 6.94$. 
Here  $\chi^2=5$ for 8 data points.}
\end{figure}

\section{Conclusion}

In the present model of the Brownian dynamics both for dynamics A and B, 
the estimate of $C=D_R/D_L$ should be consistent 
with that of synthetic polymer experiments, 
while it is smaller than that of  DNA experiments. 
The difference in the ratio $C$ between  
synthetic polymers and DNAs may be due to 
some additional potential functions arising from the closed 
DNA double strands.

\section*{Acknowledgments}
The authors would like to thank Dr. K. Tsurusaki 
for helpful discussions and valuable comments. They are also grateful to 
Mirei Takasoe for useful comments. 
The present study is partially supported 
by Grant-in-Aid for Scientific Research Tokutei Ryouiki 19031007.

\appendix

\section{Algorithm of the Brownian dynamics}
 
In the paper we have simulated linear and ring polymers in a good 
solvent with hydrodynamic interaction 
by the revised version of the Brownian dynamics \cite{Ermak} 
with respect to the first-order 
predictor-corrector \cite{Iniesta}.

Let us explain the original version of 
the Brownian dynamics \cite{Ermak}.  We consider 
$N$ Brownian particles in a solvent of temperature $T$ 
with hydrodynamic interaction.  
The position of the $i$th Brownian particle, $\vec{r}_i$,  
at time $t+\Delta t$ is calculated by the following equation:   
\begin{equation}
\Delta \vec{r}_i=\vec{r}_i(t+\Delta t)-\vec{r}_i(t)=
\sum_j \frac{\partial D_{ij}}{\partial \vec{r}_j}
+\sum_j \frac{ D_{ij} \vec{F}_j}{k_B T}+\vec{R}_i(\Delta t)
\label{eq:A1}
\end{equation}
for $i, j = 1, 2,..., N$. 
Here, $D_{ij}$ denote the diffusion tensor, 
$\vec{F}_j$ the force acting on the $j$th particle, 
which we shall specify shortly. 
$\vec{R}_i(\Delta t )$ denote random numbers 
obeying the Gaussian distribution with 
 $\langle \vec{R}_i(\Delta t) \rangle =0 $ and 
$\langle R_{i \alpha}(\Delta t) R_{j \beta}(\Delta t) \rangle= 
2D_{ij} \delta_{\alpha \beta} \Delta t$. 

We derive  (\ref{eq:A1}) as follows.  
First, we consider the Fokker-Planck equation of $N$ Brownian particles 
in a solvent 
\begin{equation}
\frac{d W}{d t}=\sum_i \sum_j \bigl(\frac{\partial}{\partial \vec{r}_i} D_{ij} \frac{\partial W}{\partial \vec{r}_j}-\frac{1}{kT}\vec{F}_j W \bigr)
\label{eq:FPeq}
\end{equation}
where $W=W(\vec{r}_1,..,\vec{r}_{N},t)$ is the distribution function 
for the configuration space of the $N$ particles. 
We can show that the distribution function  
is given by the multi-variable Gaussian 
distribution if the initial configuration of the $N$ Brownian 
particles is given by 
$W({\vec{r}_1^0,...,\vec{r}_N^0},0)=\prod_{i} \delta(\vec{r}_i-\vec{r}_i^0)$. 
Up to the first order of $\Delta t$, 
the average value and the variance-covariance of
the Gaussian distribution, respectively,  are given by 
the following:  
\begin{eqnarray}
\langle\Delta \vec{r}_i\rangle & = & \sum_j \frac{\partial}{\partial \vec{r}_i} D_{ij} (\frac{\partial W}{\partial \vec{r}_j}-\frac{1}{kT}\vec{F}_j W) \, , 
\label{eq:average} \\ 
\langle\Delta r_{i \alpha} \Delta r_{j \beta}\rangle & =& 
2D_{ij}\delta_{\alpha \beta}\Delta t \, .  \label{eq:variance}
\end{eqnarray}
We thus obtain equation (\ref{eq:A1})   
from the conditions that the difference of the position vector 
$\Delta \vec{r}_i=\vec{r}_i(t+\Delta t)-\vec{r}_i(t)$ should 
satisfy the average value (\ref{eq:average}) 
and the variance-covariance (\ref{eq:variance}). 
Here we remark that 
we can obtain the same average value (\ref{eq:average}) 
and the variance-covariance (\ref{eq:variance}) 
by integrating the Langevin equations of $N$ Brownian particles. 

Let us now formulate the diffusion tensor and the force acting 
on the Brownian particles.  
We employ the Ronte-Prager-Yamakawa tensor as the diffusion tensor 
\cite{Rotne,Yamakawa}:  
\begin{eqnarray}
D_{ij} & =& \frac{kT}{6\pi\zeta a}\delta_{ij} \quad ({\rm for} \, i=j) \\ 
D_{ij}& =& \frac{kT}{8\pi \zeta r_{ij}}
\biggl[ \bigl(E+\frac{\vec{r}_{ij}\vec{r}_{ij}}
{r_{ij}^2}  \bigr)+\frac{2a^2}{r_{ij}^2}
\bigl(\frac{1}{3}E-\frac{\vec{r}_{ij}\vec{r}_{ij}}{r_{ij}^2}   \bigr)  
 \biggr] \quad ({\rm for} \,  i \not= j)
\label{eq:Dtensor}
\end{eqnarray} 
Here $a$ denotes the radius of a bead and $\zeta$ the hydrodynamic friction. 
For the force, we assume the Lennard-Jones force and the FENE spring force. 
The Lennard-Jones potential is given by 
\begin{equation}
V_{LJ}=4 \epsilon_{\rm LJ} \bigl( \bigl( \frac{\sigma_{\rm LJ}}{r} \bigr)^{12}-\bigl( \frac{\sigma_{\rm LJ}}{r}\bigr)^6 \bigr)
\label{eq:LDpot}
\end{equation}
where $r$ is the distance between two particles, 
$\sigma_{\rm LJ}$  the zero-energy distance 
and $\epsilon_{\rm LJ}$ the energy at distance 
$\sigma$. We give $\sigma_{\rm LJ}=0.8b$ and $\epsilon_{\rm LJ}=0.1k_B T$ 
for simulation in a good solvent. 
The potential of the FENE spring force is given by 
\begin{equation}
V_{FENE}=- {\frac 1 2} r_{max}^2H \ln 
\bigl[  1-(\frac{r}{r_{max}})^2 \bigr]
\label{eq:FENE} 
\end{equation}
where $r$ is the distance between a pair of neighboring particles, 
$H$ the spring constant and $r_{max}$ the maximal distance between 
neighboring particles. 
For dynamics B, we set $H^{*}=3.0$ and $r^{*}_{max}=10.0$, which are  
given in \cite{Cifre}. 
For dynamics A, we set $H^{*}=30.0$ and $r^{*}_{max}=1.3$, as shown in  
Ref. \cite{Kremer-Grest}. In this model no bond crossing should be possible 
due to the strong spring constant and the small maximal distance between
neighboring particles.
Here we note that dimensionless parameters and variables are obtained 
by dividing length, time and energy 
 by $b$, $\zeta b^2/kT$ and $kT$, respectively.

The first-order predictor-corrector version \cite{Iniesta} of the 
Ermak and McCammon algorithm  \cite{Ermak} 
is given as follows.  When initial positions of all particles
$\vec{r}_{i}^{\, 0}$ are given, we calculate the positions at 
the next time step as follows. First,  
we calculate the diffusion tensor and the 
force, i.e. $D_{ij}^{0}$ and $\vec{F}_i^{\, 0}$, respectively,  
making use of (\ref{eq:Dtensor}), (\ref{eq:LDpot}) and (\ref{eq:FENE}). 
Second, we calculate the positions of all particles, $\vec{r}_i^{\, '}$,  
by (\ref{eq:A1}) with respect to $D_{ij}^{0}$ and $\vec{F}_i^{\, 0}$. 
Third, using $\vec{r}_i^{\, '}$, we again  calculate the 
diffusion tensor and the force, and denote them by $D_{ij}^{'}$ and
$\vec{F}_i^{\, '}$, respectively. 
Finally, we calculate the position of the $i$th particle at the next time step 
as follows. 
\begin{eqnarray}
 \Delta \vec{r}_i & = & \vec{r}_i(t+\Delta t)-\vec{r}_i^{\, 0}(t) 
 =  \Delta t \sum_{j} 
\frac{1}{2} \bigl( 
\frac{\partial}{\partial \vec{r}_{j}^{\, 0}} D_{ij}^{0} + 
\frac{\partial}{\partial \vec{r}_{j}^{\, '}} D_{ij}^{'} \bigr) 
\nonumber \\ 
& + & \Delta t \sum_{j} \frac{1}{2} \bigl(D_{ij}^{0}\vec{F}_{j}^{0}+D_{ij}^{'}
\vec{F}_{j}^{\, '}  \bigr)/k_B T + \vec{R}_{j}   
\end{eqnarray}
Here $\vec{R}_{j}$ obey the Gaussian distribution where the average value 
is zero and the variance-covariance is given by the following: 
\begin{equation}
\langle R_{i \alpha}R_{j\beta}\rangle=
2\bigl[\frac{1}{2 }\bigl(D_{ij}^{0}+D_{ij}^{'}  \bigr)
\bigr]\delta_{\alpha \beta} \Delta t. 
\end{equation} 

\end{document}